\documentclass[aps,prb,amsmath,twocolumn,amssymb,foo
tinbib,superscriptaddress]{revtex4}

\usepackage[apple mac]{inputenc}
\usepackage[english]{babel}
\usepackage{times}
\usepackage{latexsym}
\usepackage{fancyhdr}
\usepackage{verbatim}

\usepackage{graphicx}
\usepackage{epsf}
\usepackage{subfigure}
\usepackage{amsmath}
\usepackage{amssymb}
\usepackage{amsfonts}
\usepackage{bm}
\usepackage{natbib}
\usepackage{epstopdf}

\def\be{\begin{equation}}
\def\ee{\end{equation}}
\def\ba{\begin{align}}
\def\enda{\end{align}}
\def\bi{\begin{itemize}}
\def\ei{\end{itemize}}
\def\bs#1{\boldsymbol{#1}}
\def\txt#1{\textrm{#1}}

\def\nm{{\ {\rm nm}}}						
\def\micron{{\ \mu{\rm m}}}					

\def\Hz{{\ {\rm Hz}}}						
\def\kHz{{\ {\rm kHz}}}						
\def\MHz{{\ {\rm MHz}}}						
\def\GHz{{\ {\rm GHz}}}						
\def\THz{{\ {\rm THz}}}						

\def\Rb87{^{87}\rm{Rb}}						
\def\Li6{^{6}\rm{Li}}						

\newcommand{\ket}[1]{|#1\rangle}

\begin{document}


\title{Engineering Time-Reversal Invariant Topological Insulators With Ultra-Cold Atoms}


\author{N.~Goldman *}
\affiliation{Center for Nonlinear Phenomena and Complex Systems - Universit$\acute{e}$ Libre de Bruxelles (U.L.B.), Code Postal 231, Campus Plaine, B-1050 Brussels, Belgium}

\author{I.~Satija}
\affiliation{
Department of Physics, George Mason University, Fairfax, Virginia 22030, USA}
\affiliation{National Institute of Standards and Technology, Gaithersburg, Maryland 20899, USA}

\author{P.~Nikolic}
\affiliation{
Department of Physics, George Mason University, Fairfax, Virginia 22030, USA}
\affiliation{National Institute of Standards and Technology, Gaithersburg, Maryland 20899, USA}

\author{A.~Bermudez}
\affiliation{
Departamento de F\'isica Te\'orica I,
Universidad Complutense,
28040 Madrid,
Spain
}

\author{M.~A.~Martin-Delgado}
\affiliation{
Departamento de F\'isica Te\'orica I,
Universidad Complutense,
28040 Madrid,
Spain
}

\author{M.~Lewenstein}
\affiliation{
ICFO-Institut de Ci\`encies Fot\`oniques,
Parc Mediterrani de la Tecnologia,
E-08860 Castelldefels (Barcelona), Spain}
\affiliation{
ICREA - Instituciio Catalana de Recerca i Estudis Avancats, 08010
Barcelona, Spain}

\author{I.~B.~Spielman}
\affiliation{Joint Quantum Institute, National Institute of Standards and Technology, and University of Maryland, Gaithersburg, Maryland, 20899, USA}

\date{\today}

\maketitle

%
%
%
%
%
%


{\bf
Topological insulators are a broad class of unconventional materials that are insulating in the interior but conduct along the
edges. This edge transport  is topologically protected and dissipationless.
Until recently, all existing topological insulators, known as quantum Hall states, violated time-reversal symmetry.
However, the discovery of the quantum spin Hall effect demonstrated the
existence of novel topological states not rooted in time-reversal violations.
Here, we lay out an experiment to realize time-reversal  topological insulators in ultra-cold atomic gases
subjected to synthetic gauge fields in the near-field of an atom-chip. In particular, we introduce a 
feasible scheme to engineer sharp boundaries where the ``edge states"  are localized. Besides, this multi-band system  has a large parameter space exhibiting a variety of quantum phase transitions
between topological and normal insulating phases. Due to their unprecedented controllability, cold-atom systems are ideally
suited to realize topological states of matter and drive the development of topological quantum computing.

}

\section{Introduction}

The search of exotic states of matter has been one of 
the key driving elements of condensed-matter physics. 
Among these unconventional phases, the quantum Hall effect has been the paradigm of
topological phases not captured by Landau symmetry breaking.
Recently, the subject of topological insulators has emerged as
a new frontier, discovering novelties in the single-particle band theory and
providing a new impetus to the constantly evolving quantum many-body physics of
strongly correlated systems~\cite{HasanKane}. The topological classiﬁcation of insulators and superconductors 
points towards an inherent universality underlying these exotic states of 
matter~\cite{Kitaev}.

A new kind of topological insulator, based on the quantum spin Hall (QSH) effect~\cite{Kane2006}, was first theoretically predicted~\cite{Kane2005bis,Bernevig2006,Bernevig2006bis,Fu2007}, then experimentally confirmed in HgTe/CdTe quantum wells~\cite{Konig2007} and three-dimensional Bi$_2$Te$_3$~\cite{Chen2009b}. Unlike quantum Hall effects, both integer~\cite{Klitzing1980} (IQHE) and fractional~\cite{Tsui1982}, the QSH effect occurs at zero magnetic field and thus respects time-reversal (TR) symmetry. The simplest model of the QSH effect consists of non-interacting charged spin-1/2 fermions where the two spin components are described as IQHE states at equal but opposite ``magnetic fields.''
This composite system produces degenerate (Kramers) pairs of counter-propagating chiral edge states with opposite spins.
Residing in the bulk gap of the band
structure, edge modes are gapless excitations that provide the backbone for the topological classification of the insulators.
  In contrast to the usual IQHE where edge states carry electrical current with quantized conductance~\cite{Hatsugai1993}, the edge states in QSH systems carry robust spin currents. In the ideal model, the latter are associated to a  quantized spin-conductivity~\cite{Kane2005,Bernevig2006}.
  
In electronic materials QSH effects result from spin-orbit (S-O) coupling, and spin-mixing interactions destroy the quantization of the spin Hall conductivity. A simple way to determine the fate of QSH insulators upon spin mixing is to count the number of Kramers edge pairs inside the bulk gap.  Due to TR symmetry, perturbations can only mix different Kramers pairs and gapless edge states are protected when their number is \emph{odd}.  This signals a QSH insulator whose properties are topologically different from a normal band-insulator (NI) that presents an \emph{even} number of edge states~\cite{edge,edgebis}. Therefore, a $Z_2$ even/odd topological index captures the distinction between QSH and NI quantum phases. The discovery of topological QSH systems has opened the possibility to design new
spintronic devices exploiting the spin-dependent direction of edge or surface states in order to manipulate and transport spin current without dissipation.

Here, we propose an optical-lattice realization~\cite{Lewenstein2007,Bloch2008} of such quantum phases and
show that cold atoms provide a simple system where quantum spin-Hall physics can be investigated.  The origin of these QSH phases is rooted in engineered gauge fields~\cite{Lin2009,Lin2009bis} rather than S-O coupling.
We describe an original scheme to synthesize gauge fields, exempt from spontaneous emission implicit in
all previous proposals~\cite{Jaksch2003,mueller,Sorensen2004,spielman,Gerbier2009,Osterloh2005,Ruseckas2005}.
The great versatility offered by cold atom experiments allows us to reach the purest realization of the QSH effect
and to explore striking aspects of this topological state of matter.
Besides, the non-interacting limit -- the ideal realization of QSH phases -- and the stability of the topological phases can be explored by means of Feschbach resonances and engineered disorder~\cite{maciejarxiv}.

\section{The system}

\begin{figure*}[htbp]
\begin{center}
\includegraphics[width=6in]{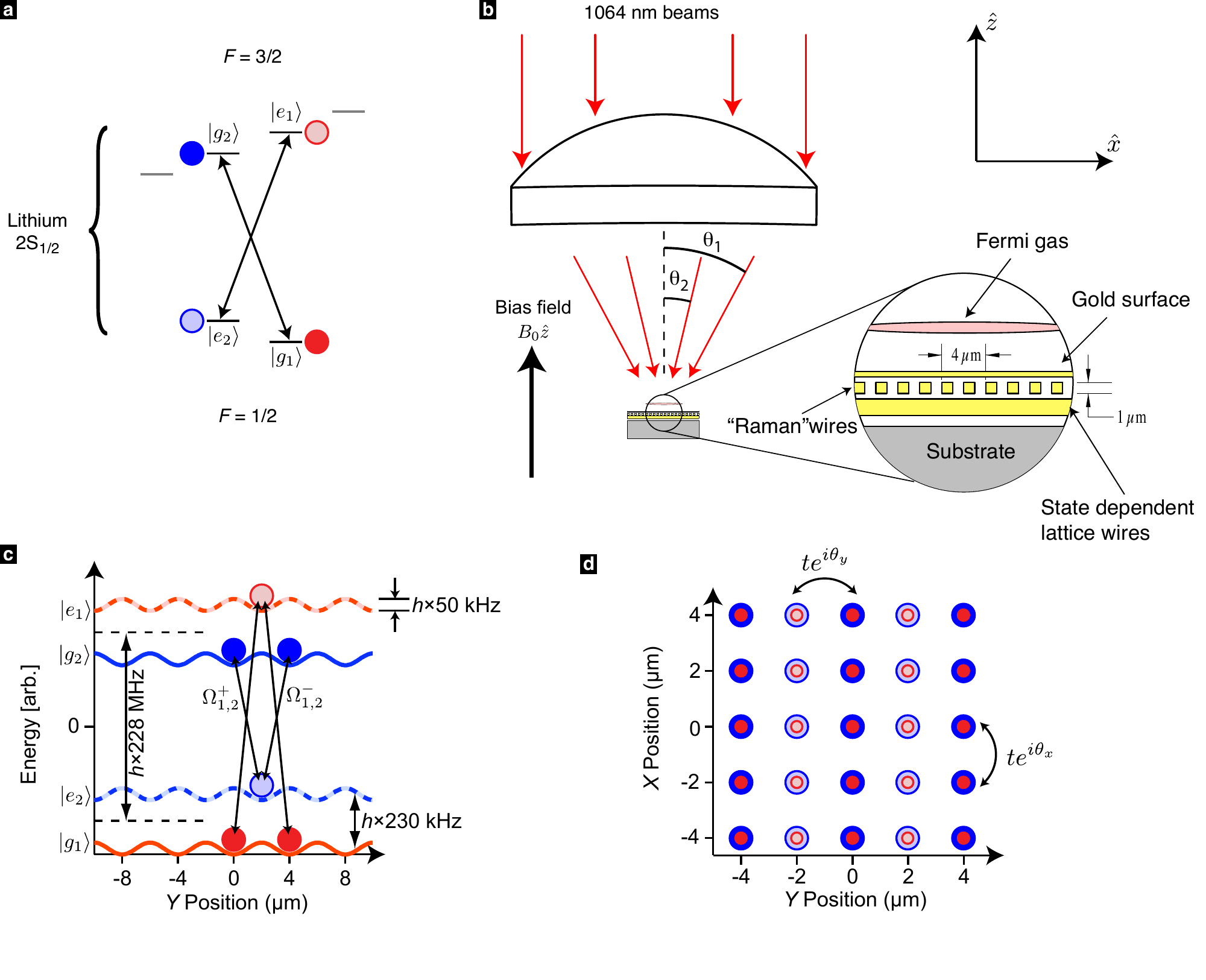}
\end{center}
\caption{{\bf Experimental realization using an atom-chip and resulting model (a)-(d):} {\bf (a)} Level diagram identifying the four required states from within the ground $2{\rm S}_{1/2}$ electronic manifold of fermionic $\Li6$.  Each of the state-pairs $\left\{\ket{g_1},\ket{g_2}\right\}$ and $\left\{\ket{e_1},\ket{e_2}\right\}$ are clock state pairs (with the same magnetic moment) and the indices $1$ and $2$ label the SU(2) degree of freedom.  Moreover the $\ket{g_{1,2}}$ and $\ket{e_{1,2}}$ pairs have opposite magnetic moments, so they experience opposite-signed magnetic potentials.  {\bf (b)} Explicit experimental layout showing the origin of optical (state independent) and magnetic (state dependent) potentials and coupling fields.  First, a state-independent, staggered, lattice along $\hat x$ is formed by the separate interference of two pairs of $\lambda=1064\nm$ laser beams slightly detuned from each other to eliminate cross interference.  The respective intersection angles are chosen so the lattice period differ by a factor of two.  Both beams reflect from the chip-surface and form vertically aligned lattices, trapping the degenerate Fermi gas about $5\micron$ above the surface.  The inset shows the chip geometry, from top to bottom: a reflective chip-surface, gold wires aligned along $\hat y$ with a $1\micron$ spacing along $\hat x$ (producing the $\ket{g}$-$\ket{e}$ coupling), and finally gold wires aligned along $\hat x$ with a $2\micron$ spacing (producing the state dependent lattice).  {\bf (c)} Computed atomic potentials and coupling matrix elements.  {\bf (d)} The resulting lattice with couplings schematically illustrated.}
\label{FIG1}
\end{figure*}

Realizing topological insulators with cold atoms is particularly attractive ~\cite{Stanescu2009,Stanescuarxiv}, and we hereby provide a concrete setup using fermionic $^{6}$Li subjected to synthetic gauge fields.  There are numerous proposals for creating Abelian~\cite{Jaksch2003,mueller,Sorensen2004,spielman,Gerbier2009} and non-Abelian~\cite{Osterloh2005,Ruseckas2005} gauge fields, leading to interesting phenomena~\cite{Goldman2009,Goldman2009bis,Satija1,Satija2,Zhu2006,oh}.  The vast majority of these proposals depend on laser-induced Raman coupling between internal atomic states; such a method was recently implemented with bosonic $^{87}$Rb atoms creating an Abelian gauge field~\cite{Lin2009,Lin2009bis}.  For the alkali atoms, the requisite coupling limits the possible detuning of the Raman laser beams from the ground S to first excited P transition, about to the excited state fine-structure splitting [$7.1\THz$ for $^{87}$Rb, $1.7\THz$ for $^{40}$K, and only $10\GHz$ for $^{6}$Li].  The very small detunings possible for fermionic atoms --  $^{40}$K and $^{6}$Li -- imply large spontaneous emission rates and thus the desired gauge fields must be synthesized through an alternate scheme.  Here we describe such a setup that combines state-independent optical potentials, with micron scale state-dependent magnetic potentials in an atom chip.  Our discussion focuses on harmonically trapped quantum degenerate $\Li6$ systems about $5\micron$ above the atom chip's surface. Figure~\ref{FIG1} depicts how the required potentials for this proposal can be realized using a combination of static and radio-frequency magnetic fields.  By completely eliminating spontaneous emission, this approach makes practical the realization of gauge fields for all alkali atoms.

Our proposal for realizing a fermionic model with a SU(2) gauge structure requires four atomic states $\ket{g_1} = \ket{F=1/2,m_F=1/2}$, $\ket{g_2} = \ket{3/2,-1/2}$, $\ket{e_1} = \ket{3/2,1/2}$, and $\ket{e_2} = \ket{1/2,-1/2}$ [see Fig.~\ref{FIG1} (a)], which are confined in a square optical lattice and described by the Hamiltonian
\begin{align}
\mathcal{H}=&-t\sum_{m,n} \bs c_{m+1,n}^{\dagger} e^{i \check\theta_{{\bf x}}} \bs c_{m,n}
+\bs c_{m,n+1}^{\dagger} e^{i \check\theta_{\bf y}} \bs c_{m,n}+\txt{h.c.}  \notag \\
&+  \lambda_{\txt{stag}} \,  \sum_{m,n}  (-1)^{m} \, \bs c^{\dagger}_{m,n} \bs c_{m,n}.
\label {2dh}
\end{align}
$\bs c_{m,n}$ is a 2-component field operator defined on a lattice site $(x= ma,y=n a)$, $a$ is the lattice spacing, $m,n$ are integers, and $t$ is the nearest-neighbor hopping.  In this model the matrices $\check\theta_{{\bf x},{\bf y}}$ modify the hopping along the $x,y$-directions and result from the synthetic gauge field [cf. below and Methods]. Here, all the states experience a primary lattice potential $V_1(x)=V_x\sin^2(k x)$ along $\hat x$ which gives rise to a hopping amplitude $t \approx 0.4$ kHz.  A secondary much weaker lattice $V_2(x)=2\lambda_{\txt{stag}}\sin^2(k x/2)$ slightly staggers the primary lattice with $\lambda_{\txt{stag}} \approx t$.  These lattices, with an approximate period of $2\micron$, are produced by two pairs of $\lambda=1064\nm$ lasers, slightly detuned from each other and incident on the reflective surface of the atom chip [see. Fig  ~\ref{FIG1} (b)]. Additionally, these beams create a lattice along $\hat z$ with a $0.55\micron$ period, confining the fermions to a 2D plane.

We propose to  design the SU(2) hopping operators as follows
\begin{align}
\check\theta_{\bf x} &= 2\pi \gamma {\bf \check\sigma_x},&&\ {\rm and}\ &\check\theta_{\bf y} &=  2\pi x \alpha {\bf \check\sigma_z},
\label{gaugefields}
\end{align}
where we have set the units $a=\hbar=1$, and ${\bf \check\sigma_{x,z}}$ are Pauli matrices. In order to engineer these state-dependent tunnelings, the states $\ket{g}$ and $\ket{e}$ must experience oppositely-signed lattices realized along the $\hat y$ direction of the chip, as in Fig.~\ref{FIG1} (c). This scheme exploits the Zeeman shift of an array of wires with alternating  currents $\pm I$. The $\check\theta_{\bf y}$ hopping operator  in Eq. \eqref{gaugefields} involves $x$-dependent phases that can be realized with an additional grid of wires spaced by $a=2\micron$ along $\hat x$ [cf. Methods]. Remarkably, these \emph{moving} Zeeman lattices reproduce the Raman-assisted hopping from previous schemes~\cite{Jaksch2003,mueller,Sorensen2004,spielman,Gerbier2009} and lead to the operator $\check\theta_{\bf y}$.  A potential gradient along $\hat y$ detunes this Raman coupling into resonance and is produced by simply shifting the center of the harmonic potential -- equivalent to adding a linear gradient.  The specific form of the hopping operator $\check\theta_{\bf y} $ is motivated by the fact that it corresponds to opposite ``magnetic fluxes" $ \pm \alpha$ for each spin component, where $\alpha=p/q$ and $p,q$ are mutually prime numbers. Finally, the additional contribution to the hopping $\check\theta_{\bf x} $ that mixes the $\ket{e}$ and $\ket{g}$ states can be realized in a like manner. The corresponding control parameter $\gamma$, and the additional staggered potential which induces alternate on-site energies $\epsilon=\pm \lambda_{\txt{stag}}$ along the $x$-axis, are shown below to enrich the phase transitions in a novel manner. Let us emphasize that the Hamiltonian  \eqref{2dh}-\eqref{gaugefields} satisfies  Time-Reversal invariance,  since it commutes with the TR-operator defined as $\mathcal{T}= i   {\bf \check\sigma_y} K$, where $K$ is the complex-conjugate operator. This  fundamental symmetry opens the possibility to realize the first  instance of $Z_2$-topological insulators in a cold-atom laboratory.  Finally, we stress that the two components of the field operators correspond in general to a pseudo-spin 1/2, but in our proposal refer specifically to spin components of $^6$Li in its electronic ground state.

In this paper, we present results for the case $\alpha=1/6$, namely when the synthetic gauge field  \eqref{gaugefields} corresponds to $\check\theta_{\bf y} (m)= 2\pi m {\bf \check\sigma_z} /6$.  This choice exhibits an extremely rich phase diagram with almost all possible topological phase transitions [cf. Figs.~\ref{FIG4}].  The topological phases discussed below are robust against small variations $\delta \alpha \sim 0.01$: indeed the properties rely on the existence of bulk gaps which are continuously deformed when $\alpha$ is varied, as in the Hofstadter butterfly~\cite{Hofstadter1976}. Moreover, other configurations of the gauge field could be experimentally designed and would lead to similar effects.  In Sect.~\ref{transition}, we investigate the phase transitions between topological insulators for a cylinder geometry in the absence of the confining potential $V_{\txt{conf}}$. We then show in Sect.~\ref{detection} how these properties can be detected when this confinement is applied.

\section{Topological insulators and phase transitions}
\label{transition}

When $\gamma=0$, Eqs.~\eqref{2dh}-\eqref{gaugefields} describe two uncoupled quantum Hall systems.  For generic $\alpha=p/q$, the fermion band-structure splits into $q$ subbands, the lattice analogues of Landau levels. These subbands are separated by gaps, or by ``pseudogaps" exhibiting nodal Dirac fermions.  Our setup thus provides a multi-band system, where a variety of band-insulators, separated by metal or semi-metal regions, can be reached by varying the atomic filling factor. As discussed below, some of these insulators are topologically non-trivial and feature gapless edge states. The latter are localized at the boundaries of the sample, and  correspond to gapless excitations.  When the Fermi energy $E_F$ lies inside a bulk gap, the presence of these states is responsible for the spin transport along the edges.

\begin{center}
\begin{figure}[h!]
\begin{center}
{\scalebox{0.2}{\includegraphics{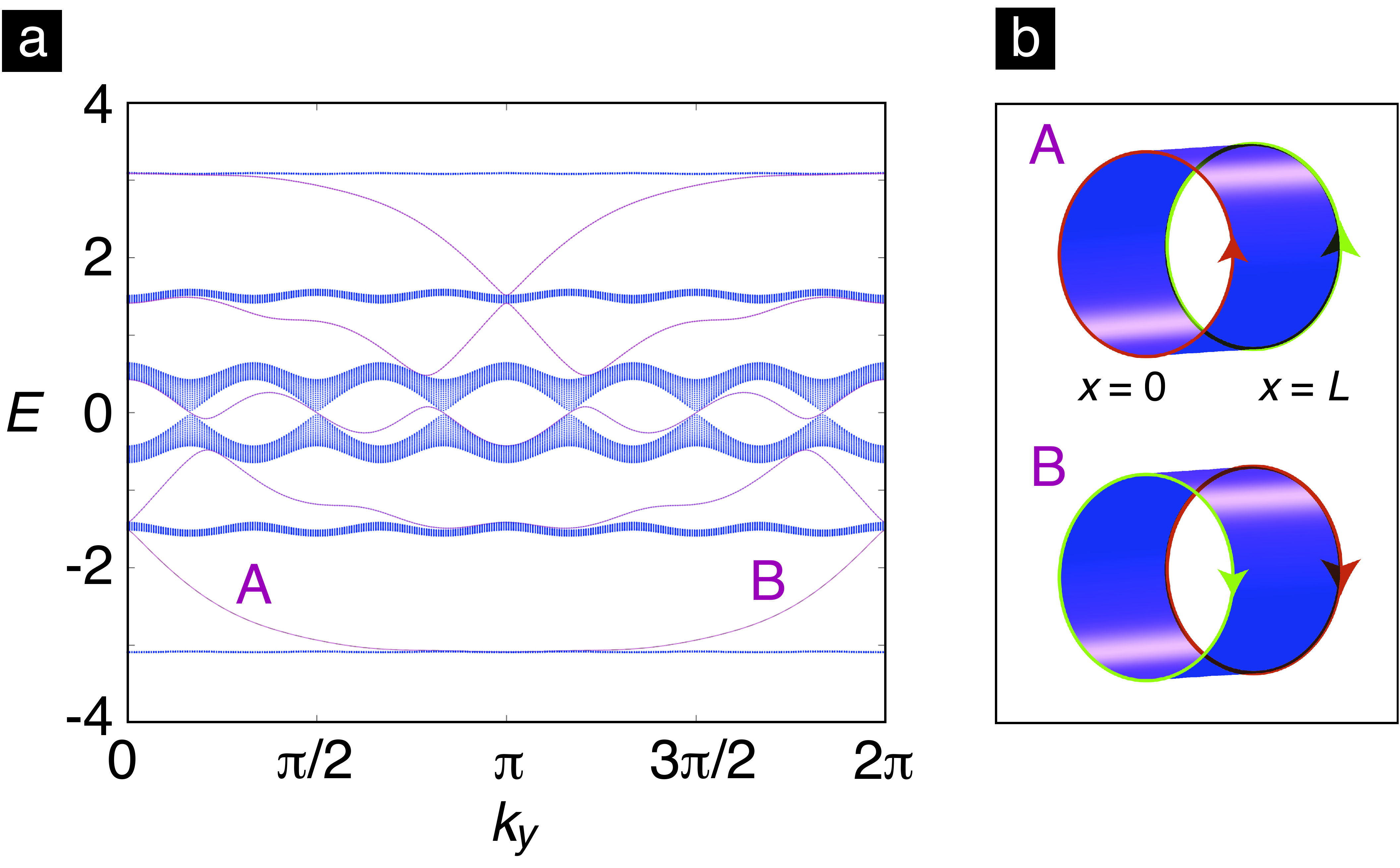}}}
\caption{\label{FIG2} {\bf Bulk energy bands and edge-state channels}: {\bf (a)}  Energy spectrum of the uncoupled system ($\gamma=0$ and $\alpha=1/6$) computed in a cylindrical geometry: the bulk energy bands [thick blue bands] are traversed by edge states [thin purple lines]. The energy is in units of the tunneling amplitude $t$. {\bf (b)} Schematic representation of the two edge states, A and B, that lie inside the first bulk energy gap depicted in (a). The \emph{spins} traveling around the edges are respectively represented by red and  green arrows.}
\end{center}
\end{figure}
\end{center}

Figure~\ref{FIG2} (a) illustrates the energy spectrum obtained for a cylindrical geometry [cf. Methods].  In this figure, the bulk bands [thick blue bands] are clearly differentiated from the gapless edge states within the bulk gaps [thin purple lines]. In the lowest bulk gap, the edge state channel A, and its TR-conjugate B, correspond to localized excitations which travel in opposite directions [cf. Fig.\ref{FIG2} (b)].  Each boundary is populated by a single pair of counter-propagating states, each corresponding to an opposite spin [i.e. helical edge states].  Accordingly, this lowest bulk gap describes a topological QSH phase.  Conversely, the next gap located at $E \approx -1$ is traversed by an even number of Kramers pairs, thus yielding a normal band insulator. Remarkably enough, different topological insulators can be experimentally engineered in this multi-band scenario by simply varying the atomic filling factor.  The Fermi surface at half-filling, namely $E_F=0$, corresponds to a discrete set of isolated points, and thus describes a semi-metal phase.

An alternative approach to the above even-odd criteria relies on the computation of a $Z_2$ topological invariant that characterizes the bulk gaps [see Methods].  The QSH phase is distinguished from the normal insulator by the $Z_2$ index $I_{Z_2}$, which respectively takes the value $I_{Z_2}(\txt{QSH})=1$ and $I_{Z_2}(\txt{NI})=0$. We computed this index inside the four gaps depicted in Fig.\ref{FIG2} (a), obtaining the following sequence
\be
I_{Z_2}= \{ 1 , 0 , 0 , 1 \} , \notag
\ee
in perfect agreement with the above edge-state analysis.  Therefore, in the uncoupled case $\gamma=0$ where spin is a good quantum number, the spin conductivity is quantized as $\sigma_s = 2  \, I_{Z_2} \, \frac{e}{4 \pi}$ and is related to the spin current $j_s=j_{\uparrow}- j_{\downarrow}$. Additionally, this quantized spin Hall conductivity is equal to the difference of the Chern numbers associated to the up- and down-spin [cf. Methods]. \\

\begin{center}
\begin{figure}[h!]
\begin{center}
{\scalebox{0.4}{\includegraphics{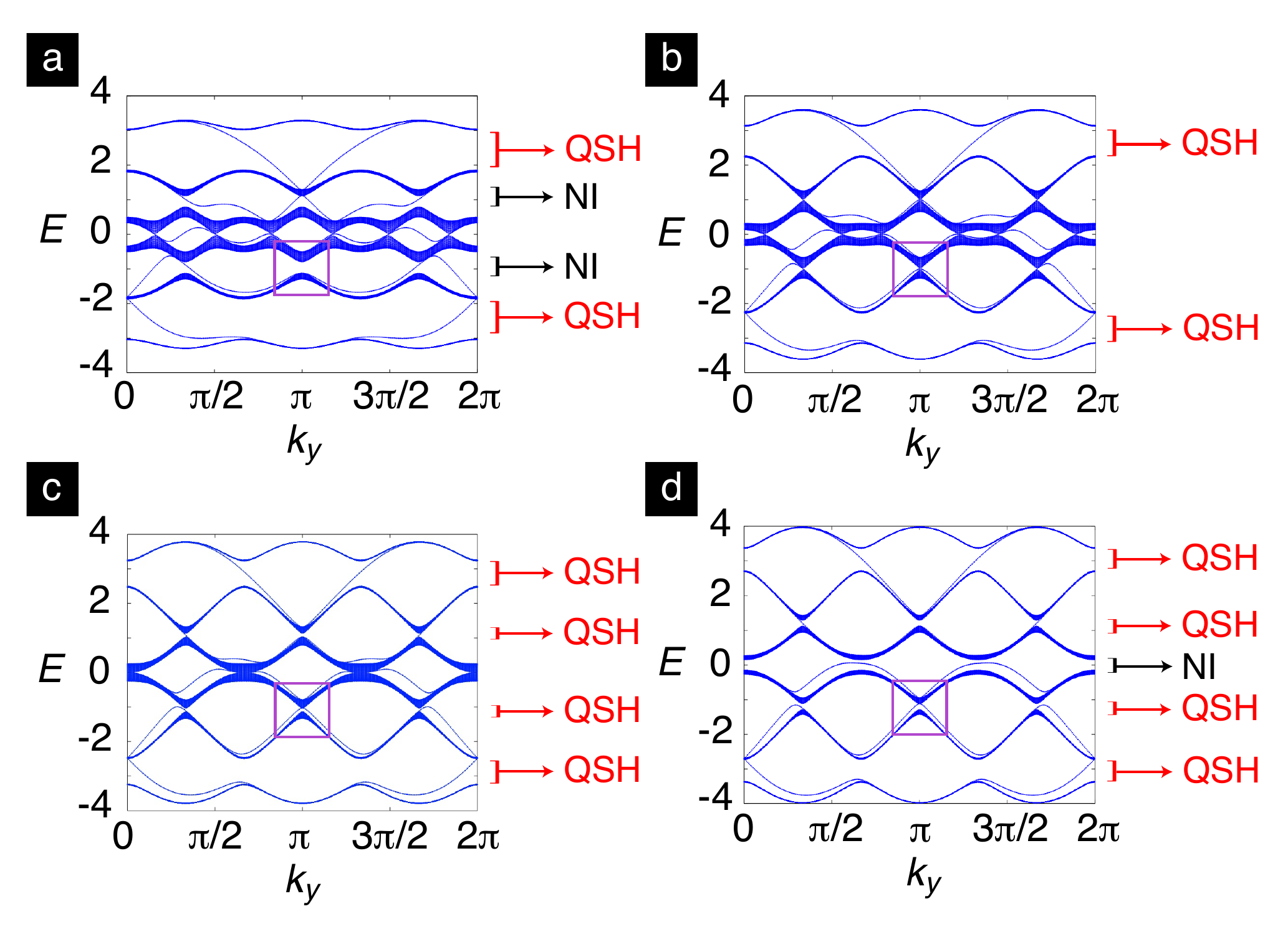}}}
\caption{\label{FIG3} {\bf Energy bands $\bs{E(k_y)}$ for $\bs{\gamma=0}$ and $\bs{\alpha=1/6}$ , in the presence of an external staggered potential (a)-(d)}: {\bf (a)} $\lambda_{\txt{stag}}=0.5 t$, {\bf (b)} $\lambda_{\txt{stag}}= t$, {\bf (c)} $\lambda_{\txt{stag}}=1.25 t$, {\bf (d)} $\lambda_{\txt{stag}}=1.5 t$. The arrows point at the open bulk gaps and indicate their corresponding topological phases. The purple rectangles highlight the NI to QSH phase transition that occurs around $E_F=1$. The energy is expressed in units of the tunneling amplitude $t$.}
\end{center}
\end{figure}
\end{center}

In our multi-band system, the lattice-potential distortions can drive direct transitions between normal and topological insulating states.  Figure~\ref{FIG3} shows the bulk gaps and edge states for increasing values of the staggered potential's strength $\lambda_{\txt{stag}}$. In the simple case $\gamma=0$, where two quantum Hall systems are superimposed, we already obtain various realizations of QSH phases. By tuning $\lambda_{\txt{stag}}$, one can observe a NI to semi-metal transition, and more importantly, a semi-metal to QSH phase transition in the  bulk gaps located around $E = \pm 1$. Remarkably enough, we have shown that the staggered potential, which can be controlled in an optical-lattice experiment, induces a quantum phase transition from a NI to a QSH phase at the critical value $\lambda_{\txt{stag}}=t$  [cf. Fig.~\ref{FIG3} (b)]. At half-filling, the existence of Dirac points appear to resist the opening of the gap for small staggered potential and eventually lead to a NI phase for $\lambda_{\txt{stag}}>1.25 t$  [cf. Fig.~\ref{FIG3} (d)]. As we describe below, the latter situation radically changes as the coupling $\gamma$ is switched on.

\begin{center}
\begin{figure}[h!]
\begin{center}
\includegraphics[width=3in]{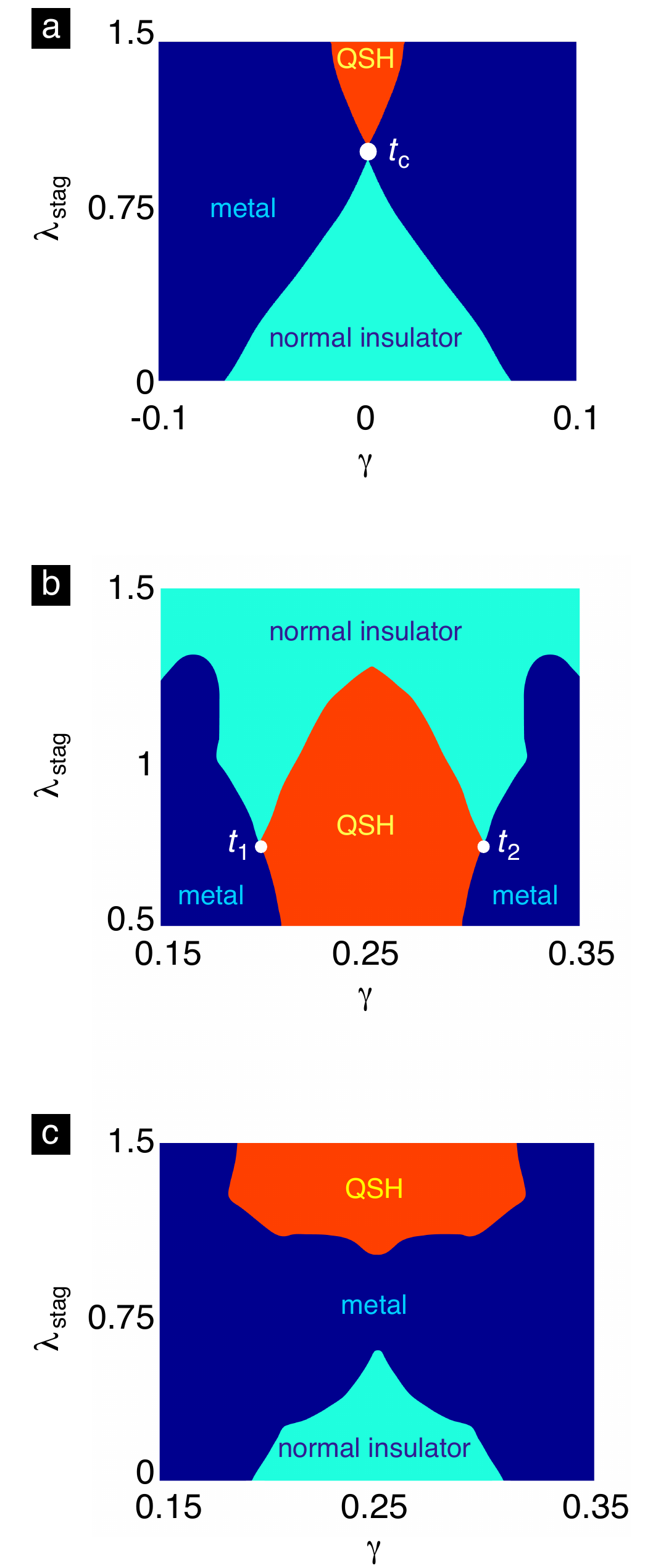}
\caption{\label{FIG4} {\bf Phase diagrams in the $\bs{(\gamma , \lambda_{\txt{stag}})}$ plane (a)-(c)}: {\bf (a)} In the vicinity of the uncoupled case $\gamma =0$ when $E_F = \pm 1$. The tricritical point $t_c$ is situated at $(\gamma =0, \lambda_{\txt{stag}}= t)$ and corresponds to Fig.~\ref{FIG3} (b). {\bf (b)} In the vicinity of the maximally coupled case $\gamma =0.25$, when $E_F = 0$: two tricritical points $t_1$ and $t_2$ are observed. {\bf (c)} In the vicinity of the maximally coupled case $\gamma =0.25$, when $E_F = \pm 1$. The red, blue and cyan regions respectively show the QSH, metallic and normal insulating phases. The staggered potential's strength $\lambda_{\txt{stag}}$ is expressed in units of the tunneling amplitude $t$.}
\end{center}
\end{figure}
\end{center}

The parameter $\gamma \ne 0$ perturbs the ideal quantum spin Hall system and fundamentally changes the aforementioned phase transitions. The $Z_2$ invariant analysis provides a particularly useful and efficient tool to obtain the full phase diagram of the system in the $(\gamma , \lambda_{\txt{stag}})$ parameters plane [cf. Methods]. The phase diagram represented in Fig.~\ref{FIG4}(a)  has been obtained numerically by evaluating the $Z_2$ index inside the bulk at $E \approx \pm 1$ in the vicinity of $\gamma=0$. Here we observe three distinct regions, namely the metallic (blue), QSH (red), and  NI (cyan) phases.  These three phases coexist at a tricritical point situated at $\gamma=0$ and $\lambda_{\txt{stag}}=t$.  It is interesting to highlight that the QSH phase occurs for a wide range of the parameter $\gamma$, which indicates the robustness of this topological phase under small local perturbations of the Hamiltonian~\eqref{2dh}.

\begin{center}
\begin{figure}[h!]
\begin{center}
{\scalebox{0.4}{\includegraphics{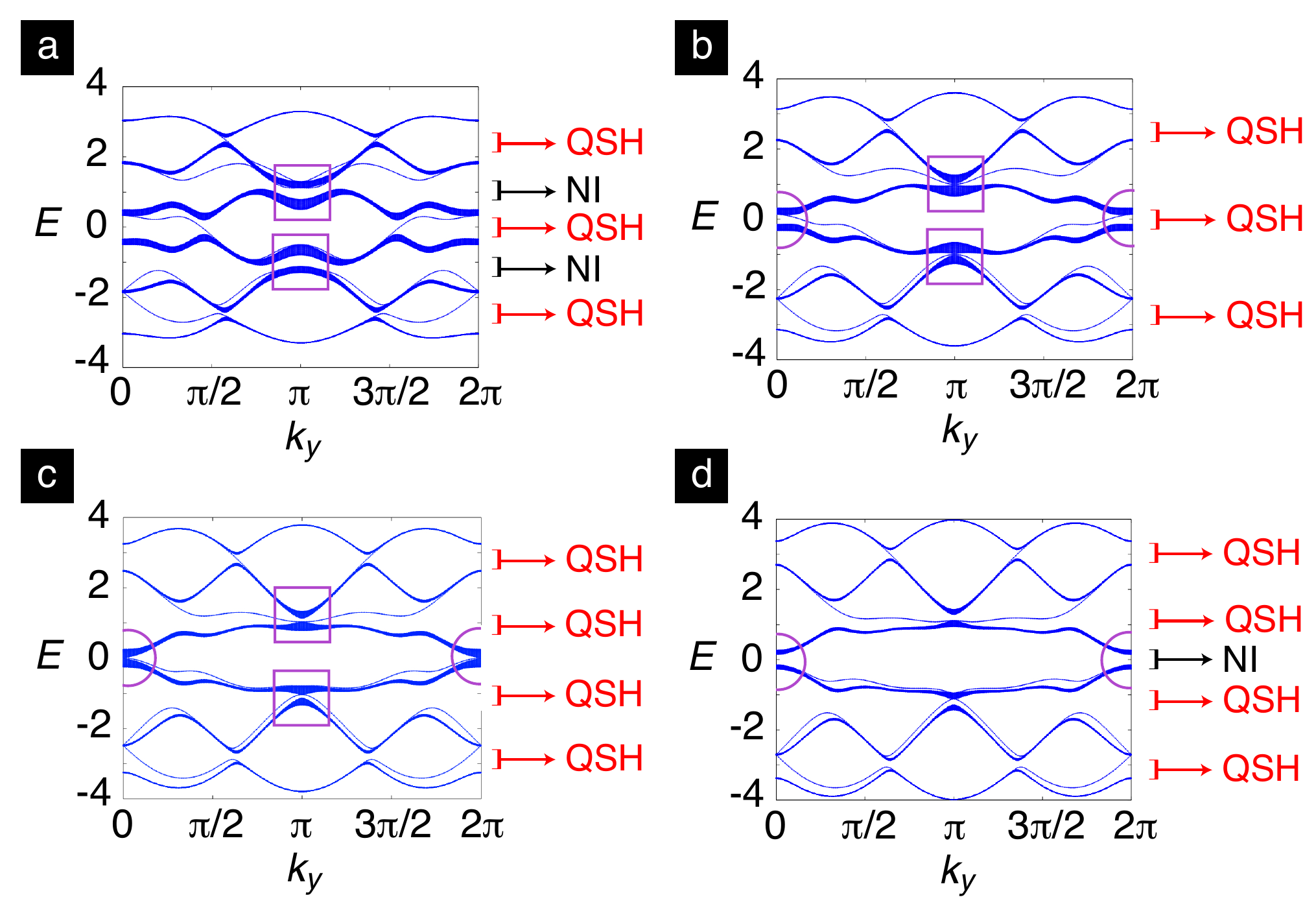}}}
\caption{\label{FIG5} {\bf Energy bands $\bs{E(k_y)}$ for $\bs{\gamma=0.25}$ and $\bs{\alpha=1/6}$, in the presence of an external staggered potential (a)-(d)}: {\bf (a)} $\lambda_{\txt{stag}}=0.5 t$, {\bf (b)} $\lambda_{\txt{stag}}= t$, {\bf (c)} $\lambda_{\txt{stag}}=1.25 t$, {\bf (d)} $\lambda_{\txt{stag}}=1.5 t$. The arrows point at the open bulk gaps and indicate their corresponding topological phases. The purple rectangles [resp. circles] highlight the NI to QSH [resp. QSH to NI] phase transitions. The energy is expressed in units of the tunneling amplitude $t$.}
\end{center}
\end{figure}
\end{center}

The wide range of possibilities offered by optical-lattice experiments enables us to consider the strong coupling regime corresponding to $\gamma = 0.25$. In this limit, the previously independent quantum Hall subsystems ($\gamma=0$) become maximally coupled, a fact that drastically modifies the topological phase transitions presented above.  In Fig.~\ref{FIG5}, we illustrate the bulk bands and edge states for increasing values of the staggered potential. Each gap is associated to a QSH or a NI phase according to the even-odd number of TR pairs. In this strong-coupled regime, we observe a completely different phase transition associated to the bulk gap at half filling. Contrary to the semi-metal to NI phase transition described in the uncoupled case [cf. Figs.~\ref{FIG3}(a)-(d)], we identify here a QSH to NI transition which occurs after a gap-closing at $ \lambda_{\txt{stag}}=1.25 t$. This process is emphasized on the figures with purple circles [cf. Figs.~\ref{FIG5} (b)-(d)].  The QSH phase can only occur at half-filling for large couplings $\gamma$. Furthermore, one also notices that the neighboring bulk gaps at $E \approx \pm 1$ present exactly the opposite phase transition, from a NI to a QSH phase. The latter occurs at $ \lambda_{\txt{stag}}= t$, as emphasized on the successive figures with purple rectangles [cf. Figs.~\ref{FIG5} (a)-(c)]. Both Figs.~\ref{FIG3} and~\ref{FIG5} show that the parameters $\gamma$ and $\lambda_{\text{stag}}$
play competing roles: $\gamma$ tends to shrink the gaps width, while $\lambda_{\text{stag}}$  tends to open them.  In this multi-band configuration, an opening of one of the gaps may result in a gap-closing of the neighboring bands.  Therefore, by manipulating the coupling between the states, the staggered potential and the Fermi energy, the system discussed here offers the possibility to explore different topological phase transitions within the several bulk gaps. These novel features  endow the topological phase diagram with an intrinsic richness and complexity, not present in other condensed-matter realizations of the QSH effect.

In our model, topological phase transitions of different nature can exist in each of the bulk gaps. As shown in Fig.~\ref{FIG5}, neighboring bulk gaps can undergo reverse transitions successively.  To fully capture the richness of this phenomenon, we numerically compute the $Z_2$ index for a wide range of the parameters around $\gamma=0.25$.  Figure~\ref{FIG4}(b)-(c) show the bulk gaps at $E=0$ and $E \approx \pm 1$, respectively. In Fig.~\ref{FIG4}(b), the central gap features tri-critical points, and the QSH-NI phase transitions occur along a well-defined curve. On the other hand, in the neighboring gap [Fig.~\ref{FIG4}(c)], the QSH to NI phase transition is separated by an intermediate metallic region.  Therefore, we conclude that a single realization of this model allows us to study a wide variety of different topological phase transitions.

\section{Detection in the presence of the confining potential}
\label{detection}

\subsection{The harmonic trap and gauge fields with anisotropic hopping}

We now describe a new feasible scheme to engineer a sharp interface where edge states can be localized. This is essential for detecting
topological states in optical lattices, where the indispensable harmonic trap used to confine atoms strongly influences the edge states
\cite{Stanescu2009,Stanescuarxiv} when
$V_{\txt{conf}} (\txt{edge}) \sim \Delta$, where $\Delta$ is the gap's width.  In the present context $\Delta \sim 0.5 -1 t$, requiring temperatures $T < 10$ nK.


\begin{center}
\begin{figure}
\begin{center}
{\scalebox{0.59}{\includegraphics{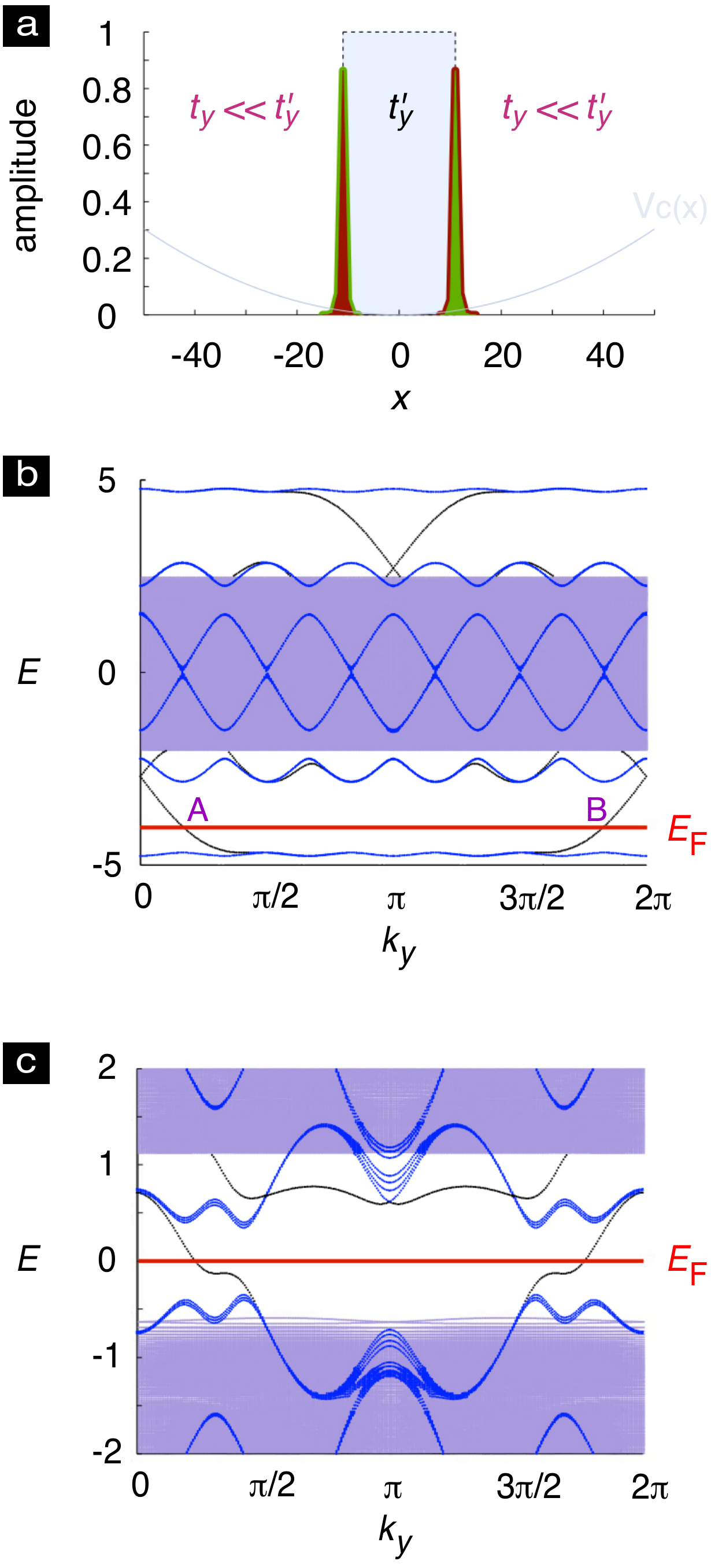}}}
\caption{\label{FIG6} {\bf Edge-states for the anisotropic-hopping scheme in the presence of the trap (a)-(c)}: {\bf (a)} The chip is divided into three regions: two sharp interfaces separate the inner region with hopping amplitude $t'_y$ from the outer regions [with hopping amplitude $t_y \ll t'_y$]. The wave-functions amplitudes $\vert \psi_{\uparrow, \downarrow} (x)\vert^2$ are localized at the interfaces and correspond to the eigenstates at $E=E_{\txt{F}}$ [cf. Fig.~\ref{FIG6} (b)].  The red [resp. green] amplitudes  correspond to up- [resp. down-] spins, and they are filled [resp. empty] according to their orientation along the ``edges". This color code highlights the helical nature of the ``edge-states" at the interfaces. {\bf (b)} Energy spectrum $E=E(k_y)$ for the case $\gamma=\lambda_{\txt{stag}}=0$ in the presence of a harmonic potential, $V_{\txt{conf}} (\txt{edge})= 0.5 t$. The hopping parameters of the inner and outer regions are respectively $t'_y=2 t$ and $t_y=0.1 t$. The eigenvalues are colored in black, blue or purple when their corresponding eigenstates are localized at the interface, the inner region or the outer regions respectively. {\bf (c)} Energy spectrum $E=E(k_y)$ for the case $\gamma=0.25$, $\lambda_{\txt{stag}}=1.125 t$ in the presence of a harmonic potential, $V_{\txt{conf}} (\txt{edge})= 0.5 t$. The hopping parameters of the inner and outer regions are respectively $t'_y=1.4 t$ and $t_y=0.1 t$. Note how the edge-states channels are robust when the anisotropic-hopping scheme is applied to the trapped system.}
\end{center}
\end{figure}
\end{center}

The key aspect of our proposal exploits the fact that the hopping along the $y$-direction, $t_y$, is laser-controlled and hence can be tuned.
Furthermore, since the harmonic trap has a minimal effect at the center of the trap,
we divide the chip into three regions: the central region is characterized by a large hopping $t'_y$, while the two surrounding regions feature small hoppings $t_y\ll t_y'$. This highly anisotropic hopping
creates a sharp interface where the edge states of the central region -- a topological insulator -- localize. In other words, by controlling the strength of the Raman coupling, we squeeze the energy bands describing the outer parts
so that it does not interfere with the bulk gaps of the central topological phase where the edge states reside. We show that the helical edge states are robust and exactly localized at the \emph{interface} between the inner and outer regions,
whose location can be easily controlled in this scheme [cf. Fig.~\ref{FIG6} (a)]. Furthermore, since the topological phases are confined in the center of the trap,
one verifies that the phase diagrams discussed in the previous sections are valid for a much wider range of the harmonic potential's strength. These observations are illustrated in Fig.~\ref{FIG6}, where the energy spectrum and edge-state wave-functions are depicted for different configurations. In Fig.~\ref{FIG6} (b), we observe a new metallic continuum band corresponding to the outer region, which corrupts all but the lowest and highest band gaps. When the Fermi energy lies in the preserved gap, we see how the edge states are robustly localized within the interface [cf. Fig.~\ref{FIG6} (a)]. Moreover, Fig.~\ref{FIG6} (c) shows that the staggered potential opens a bulk gap at half-filling, allowing to explore the phase diagrams of Fig.~\ref{FIG4} (b)-(c).

\subsection{Detecting the edge states and the quantum spin Hall effect}

Probing the topological edge states in a cold-atom experiment remains a fundamental challenge~\cite{Stanescu2009,Stanescuarxiv,Liu2010}.  Although density measurements might already detect the accumulation of atoms at the designed edges [i.e., the aforementioned interfaces], alternate techniques should access more profound properties associated to these specific states. For instance, stimulated Raman spectroscopy techniques can measure the dispersion relation of these excitations~\cite{Dao2007}. A further promising possibility would be to exploit light Bragg scattering in order to observe signatures in the dynamical structure factor~\cite{Stamper1999}: whenever the Fermi energy crosses a gapless edge-state, a delta peak is distinctively observed in the structure factor~\cite{Liu2010}. The present proposal is particular suited for testing the latter detection method, since the light scattering can be focused on the interfaces where the edge-states have been shown to be robust in the presence of a confining harmonic potential.

Finally, supplementary signatures of the quantum spin Hall effect could be detected through spin-resolved density measurements~\cite{Zhu2006}.  A synthetic electric field [e.g. by accelerating the lattice or using gravity], should separate the different spin-components, thus detecting spin-currents in the transverse direction. In the $\gamma=0$ case, the spin-Chern number is given by SChN=$N_{\uparrow}-N_{\downarrow}=2 N_{\uparrow}$ [see Methods], which can be evaluated through the Streda formula applied to the spin-up density~\cite{Umucalilar2008}.

\section{ Conclusions and Outlook}

In this article, we described a concrete and promising proposal of synthetic gauge fields in square optical lattices that overcomes the severe drawbacks affecting earlier schemes. Moreover, we showed how such engineered gauge fields are ideally suited to experimentally realize the most transparent QSH phase,  simplifying the usual S-O induced effects in condensed-matter systems.  Accordingly, ultra-cold atoms offer a rich playground to study time-reversal topological states of matter.

In systems where SU(2) gauge fields encode a non-trivial topology in the band structure, we showed how a staggered potential can drive a wide range of quantum phase transitions between metallic, QSH and NI phases.
Both NI-QSH and  QSH-NI transitions occur successively in different bulk gaps for different values
of the staggered potential.  Such multi-band gap-dependent quantum phase transitions constitute a unique and rich feature of our system.

The cold-atom realization of topological band-insulators and helical metals proposed in this paper will pave the way for engineering correlated topological superfluids and insulators. Superfluids arising from pairing instabilities in helical metals are of particular importance for quantum computing due to the existence of Majorana fermions in vortex cores~\cite{biboutlook, biboutlook2}. Highly controllable cold atoms are not only better suited than condensed-matter systems for manipulating vortices and quantum information stored in their Majorana modes, but also have slow dynamics, making them very promising for quantum computing applications.

Finally, let us  comment on the possibility of synthesizing SU(N) gauge fields with $N > 2$. Considering fermionic atoms with higher hyperfine levels $F>\frac{1}{2}$, it is possible to create artificial gauge fields where $N=2F+1$ is even. In this respect, one can envisage situations where the helical edge states present a richer spin structure, and thus offer the opportunity to explore new avenues and exotic topological phases.

\section{Methods}

\subsection{Synthetic gauge fields and the quantum spin Hall effect}

In condensed-matter systems, the quantum spin Hall effect finds its origin in the inherent spin-orbit coupling of the materials~\cite{Kane2005bis,Bernevig2006}. In the seminal lattice model of Ref.~\cite{Kane2005bis}, this term involves next-nearest-neighbor (NNN) hoppings on a honeycomb-lattice
\begin{equation}
\mathcal{H}_{\txt{S0}}= i \lambda_{\txt{S0}} \sum_{\langle \langle i,j \rangle \rangle} \nu_{ij} c_i^{\dagger} \check\sigma_z c_j ,
\end{equation}
where $\nu_{ij}=\pm1$ according to the orientation of the path separating the NNN $i$ and $j$. In analogy to the Haldane model for spinless electrons~\cite{Haldane1988}, this term can be expressed as
\begin{equation}
\mathcal{H}_{\txt{S0}}= \lambda_{\txt{S0}} \sum_{\langle i,j \rangle^{\triangle}} c_i^{\dagger} e^{i \int_j^i \bs{\mathcal{A}}\cdot \bf{dl}} c_j ,
\end{equation}
where $\bs{\mathcal{A}}=\bf{A} \check\sigma_z$ is a gauge field defined on the two triangular sub-lattices, and such that $\int_j^i \bs{\mathcal{A}} \cdot \bf{dl} = \pm \frac{\pi}{2}  \check\sigma_z$ according to the orientation of the path. Note that on a lattice, the gauge field $\bs{\mathcal{A}}$ modifies the hopping of a particle through the standard Peierls substitution\cite{Hofstadter1976}. In the original model of Haldane [i.e. $\bs{\mathcal{A}}=\bf{A}$ and $\int_j^i \bs{\mathcal{A}} \cdot \bf{dl} = \pm \frac{\pi}{2} $], this term opens an integer quantum Hall gap, while in the spin-1/2  model of Ref.~\cite{Kane2005bis}, this term is proportional to the Pauli matrix $\check\sigma_z$ and thus leads to a quantum spin Hall gap. Similarly, in the GaAs semi-conductor system of Ref.~\cite{Bernevig2006}, the spin-orbit coupling is described by a gauge field $\bs{\mathcal{A}}=\bf{A} \check\sigma_z$, where $\bs{\nabla} \times \bf{A}$ is analogous to a uniform magnetic field. In both models, the presence of the Pauli matrix $\check\sigma_z$ in the gauge field $\bs{\mathcal{A}}$ highlights the fact that  quantum spin Hall systems can be interpreted as the coupling of two quantum Hall subsystems.

In the present setup, one proposes to engineer the coupling of two Hofstadter subsystems. The Hofstadter system describes spinless electrons on a square lattice subjected to a uniform magnetic field and leads to a multi-gap energy spectrum characterized by non-trivial topological invariants~\cite{Thouless1982}. Accordingly, our square lattice Hamiltonian \eqref{2dh} features the hopping matrices $\check\theta_{\bf x,y} $, which are controlled in our optical-lattice setup, and which can be associated to the presence of a synthetic gauge field $\bs{\mathcal{A}}$ as follows
\be
\check\theta_{\bf x} = \int_{_{(m,n)}}^{^{(m+1,n)}}\bs{\mathcal{A}} \cdot \txt{d} {\bf l} \,  , \hspace{2ex} \check\theta_{\bf y} = \int_{_{(m,n)}}^{^{(m,n+1)}}\bs{\mathcal{A}} \cdot \txt{d} {\bf l} \, ,
\ee
where we have set $\hbar=c=e=1$.  Here, the synthetic gauge field $\bs{\mathcal{A}}$ is chosen in order to induce the  $\check\theta_{\bf x,y}$ operators introduced in Eq. \eqref{gaugefields}. The specific form of the hopping operator $\check\theta_{\bf y} \propto \alpha \check\sigma_z$  corresponds to opposite ``magnetic fluxes" $\pm \alpha$ for each spin component and thus leads to quantum spin Hall gaps.

\subsection{Experimental proposal}

Here we describe our technique to create an atomic ensemble subject to a SU(2) gauge field (focusing on $\Li6$, but equally applicable to $^{40}\rm{K}$) that combines state-independent optical potentials, with micron scale state-dependent magnetic potentials near the surface of an atom chip.  Our discussion focuses on harmonically trapped quantum degenerate $\Li6$ systems a distance $h=5\micron$ above an atom chips surface, in the presence of lattices with period $a=2\micron$ along $\hat x$ and $\hat y$.  We use $\hbar k_L=\pi\hbar/a$ and $E_L=\hbar^2 k_L^2/2m=h\times 8.3\kHz$ -- the recoil momentum and energy momentum -- as the natural dimensions of energy and momentum.

All proposals to create light-induced gauge fields for ultracold neutral atoms somehow involve Raman transitions that couple ground states of unlike spin that differ in momentum by $\hbar{\bf \delta k}$ by way of intermediary (and in practice off-resonant) excited states.  For a typical system with two ground states, the resulting effective Hamiltonian has (potentially many) terms of the form \begin{align*}
H'({\bf x})&= \frac{\Omega}{2}\left(\begin{array}{cc} 0 & \exp\left[i ({\bf \delta k}\cdot \hat {\bf x}+\phi)\right] \\ \exp\left[-i ({\bf \delta k}\cdot \hat {\bf x}+\phi)\right] & 0 \end{array}\right)\\
 &= \frac{\Omega}{2} \left[\cos\left({\bf \delta k}\cdot \hat {\bf x}+\phi\right)\check\sigma_{\bf x} - \sin\left({\bf \delta k}\cdot \hat {\bf x}+\phi\right)\check\sigma_{\bf y}\right].
\end{align*}
Formally this is the same as the linear Zeeman shift $g\vec\mu\cdot\vec B$ for a magnetic field in the $x$-$y$ plane orbiting about $\hat z$ with spatial period $2\pi/|{\bf \delta k}|$.  We take advantage of this formal similarity and constructs the requisite magnetic field near the surface of an atom chip.


Our proposal for realizing an atomic model with a SU(2) gauge structure requires four atomic states with very specific properties outlined below. These physically correspond to hyperfine states, labeled by $\ket{F,m_f}$, in the $\Li6$ 2S electronic ground state: $\ket{g_1}=\ket{1/2,1/2}$,$\ket{g_2}=\ket{3/2,-1/2}$,$\ket{e_1}=\ket{3/2,1/2}$, and $\ket{e_2}=\ket{1/2,-1/2}$. For simplicity, the labels $\ket{g}$ and $\ket{e}$, have been preserved from Ref.~\cite{Osterloh2005} where they denoted physical ground and excited states. 

(A) The atomic states must all experience a primary lattice potential $V_1(x)=V_x\sin^2(k x)$ along $\hat x$; a lattice depth $V_x=6 E_L\approx h\times50\kHz$ produces a hopping matrix element $t\approx0.05 E_L \approx h\times400\Hz$.  A secondary, much weaker lattice $V_2(x)=2\lambda_{\txt{stag}}\sin^2(k x/2)$ slightly staggers the primary lattice, where $\lambda_{\txt{stag}} \sim t\ll V_x$.  Figure ~\ref{FIG1} (b) shows two pairs of $\lambda=1064\nm$ lasers, slightly detuned from each other, incident on the atom chip's reflective surface, producing lattices with period $\lambda/2\sin\theta$.  For our base $a=2\micron$ lattice period, these angles [shown in Fig.~\ref{FIG1}] are $\theta_1\approx15^\circ$ and $\theta_2\approx7.5^\circ$.  These beams also produce a lattice along $\hat z$ with a $550\nm$ period, confining the fermions to a 2D plane.


(B) The states $\ket{g}$ and $\ket{e}$ experience oppositely-signed lattice potentials along $\hat y$ as in Fig.~\ref{FIG1} (c).  This can be directly implemented with the Zeeman shift $g\mu_B \left|B\right|$ of atoms provided that $\ket{g_{1,2}}$ and $\ket{e_{1,2}}$ have equal, but opposite magnetic moments $g$ (i.e., are each on clock transitions).  For the states -- $\ket{g_1}$, $\ket{g_2}$, $\ket{e_1}$, and $\ket{e_2}$ -- shown in Fig.~\ref{FIG1} (a) the magnetic moments are correctly signed and differ by less than 1\% in magnitude at a bias field $B=0.25\ {\rm G}$.

With these states, a state-dependent lattice potential can be generated by an array of current carrying wires with alternating $+I$ and $-I$ currents, spaced by a distance $a$ (repeating with full period $2a=4\micron$); because the evanescent magnetic field decays exponentially $\propto\exp\left(-2\pi h/a\right)$, the height $h$ of the atoms above the surface must be comparable to $a$.  A modest $I=5\ \mu{\rm A}$ current~\cite{Trinker2008} in wires $3\micron$ below the chip-surface ($8\micron$ from the atoms) produces a $6\ E_L$ Zeeman lattice, with a negligible $3\Hz$ hopping matrix element.

(C)  Our model also requires hopping along $\hat y$ with an $x$ dependent phase:  $t_y \exp(i q x)$.  This can be realized with an additional grid of wires spaced by $a=2\micron$ along $\hat x$, with currents $I_m$. This provides {\it moving} Zeeman lattices with wave-vector $q$ (leading to effective ``Raman coupling'') when $I_n = I_0 \sin(q m a  - \omega t)$.  The $\omega/2\pi \approx 228\pm0.23\MHz$ transitions indicated with arrows in Fig.~\ref{FIG1} (c), $\ket{e_1}\rightarrow\ket{g_1}$ and $\ket{e_2}\rightarrow\ket{g_2}$, are independently controllable in phase, amplitude, and wave-vector by commanding concurrent running waves at the indicated resonant frequencies.  For this study we confine ourselves to the equal amplitude and wave-vector case.  The minimum wavelength $d=2\pi/q$ of this moving lattice is Nyquist limited by $d > 2a$.

In the frame rotating at the angular frequency $\omega$, and after making the rotating wave approximation the coupling terms have the desired form $t \exp(i q m a)$.   In the model Hamiltonian~\eqref{2dh}, the phase for hopping along $\hat y$ is $\alpha=q a/2\pi = a/d$ in terms of physical parameters.  In this proposal we focused on $\alpha=1/6$, implying $d=6a$ far exceeding the $2a$ Nyquist limit.

Our scheme also requires, a contribution to the hopping along $\hat x$ that mixes the $\ket{e}$ and $\ket{g}$.  Like in (C) this can be realized using a Zeeman lattice moving along $\hat x$, but tuned to drive transitions between $\ket{g_1}\rightarrow\ket{g_2}$ and $\ket{e_1}\rightarrow\ket{e_2}$.


(D) To create effective edges within the trapped cloud of atoms, we propose to abruptly change $t_y$ from a larger value in the bulk to a smaller value $t_y'$ as a function of $x$.  Here, creating such a change in $t_y$ can be simply implemented by making the currents $I_0$ in our wires $x$-dependent such that they sharply change from a large value in the center to a smaller value near the systems edge, as is schematically shown in Fig.~6 (a).

(E) Lastly, a potential gradient along $\hat y$ detunes into resonance the Raman fields providing the $\hat y$ tunneling and is produced by simply shifting the center of the harmonic potential -- equivalent to adding a linear gradient.  This gradient implies that the coupling fields $\omega^{\pm}$ will be slightly different in frequency to be simultaneously resonant.  We select a gradient so the site-to-site (spaced by $a$) detuning is large compared to both the $\approx3\Hz$ bare and $\approx400\Hz$ Raman stimulated hopping along $\hat y$, but small compared to the $\approx20\kHz$ band spacing along $\hat y$.  For example a $1\kHz/a = 500\Hz/\mu{\rm m}$ gradient results from shifting a typical $40\Hz$ harmonic trap by $500\micron$.  This can be effected more efficiently in the laboratory by using the linear gradient at the side of gaussian laser beams. 


\subsection{The cylindrical geometry description}

In order to compute the edge states, we consider a cylindrical geometry with boundaries at $x=0$ and $x=L$. In this case, we apply periodic boundary conditions along the $y$-direction, and consider the partial Fourier transform $\bs c_{m,n}=\sum_{k_y}e^{-i k_y n}\bs c_{k_y}(m)$. The  Hamiltonian \eqref{2dh} becomes a collection of independent tight-binding models along the $x$-axis $H=\sum_{k_y}H_{\text{TB}}(k_y) $, where
\begin{align}
H_{\text{TB}}(k_y)=&-t \sum_m \bs c_{k_y}(m+1)^{\dagger}e^{i \check\theta_{\bf x} }\bs c_{k_y}(m)+\text{h.c} \notag\\
&+\sum_m\bs c_{k_y}(m)^{\dagger} D(m) \bs c_{k_y}(m)  ,
 \end{align}
and where the diagonal matrix yields
\begin{align}
D(m)=&-2 t \, \txt{diag} \bigl (\cos( 2\pi \alpha m+k_y) ,\cos( 2\pi \alpha m-k_y) \bigr ) \notag \\
&+\lambda_{\txt{stag}}  (-1)^m \, \mathbb{I}. \label{Dmat}
\end{align}

The energy spectra computed with this method presents the bulk bands as well as the edge states lying inside the bulk energy gaps [cf. Fig.~\ref{FIG2}].

\subsection{The $Z_2$ index and the spin-Chern numbers}

In analogy to the topological characterization of the integer quantum Hall effect by means of Chern numbers~\cite{Thouless1982,Niu1985,Kohmoto1985}, we evaluate the $Z_2$ invariants of the QSH effect through the computation of the ``\emph{spin-Chern-number}"(SChN) ~\cite{Sheng2006,Qi2006, Fukui2007}. The latter topological invariant is associated to the bulk and is therefore defined on the torus obtained by closing the cylindrical geometry [i.e. by identifying the two edges at $x=0$ and $x=L$]. Remarkably, the non-triviality of the SChN witnesses the presence of gapless edge-states, hence building an elegant bridge  between the open and closed geometry descriptions ~\cite{Qi2006}.

The computation of the SChN requires the consideration of twisted-periodic-boundary conditions \cite{Niu1985,Sheng2006,Qi2006}: as a particle crosses the boundary located at $x=0=L$, it acquires an extra phase $e^{i \check\theta_{\alpha}}$, where $\alpha=\uparrow, \downarrow$ is the spin index. Since the spectral bulk gaps and their correspondong topological order are immune to the boundary conditions, the topological SChN is obtained as the average of the Berry's curvature over all the phases aquired through the boundary. Here we set $\theta_{\uparrow}=-\theta_{\downarrow}=q \theta$ for conveniency \cite{Sheng2006, Fukui2007}, and the twist-space spanned by $(\theta, k_y)$ describes a torus: $\theta \in [0, \frac{2 \pi}{q}]$ and $k_y \in [0, 2 \pi ]$. Note that these boundary conditions break the translational symmetry along the $x$-direction, while preserving it along the $y$-direction. Besides, the closing of the cylinder is consistent as long as $L= \txt{integer} \times q$.  The single-particle Schr\"odinger equation can then be expressed in the form of a  Harper-like difference equation
 \begin{align}
&R \, \psi_{m+1} + R^{\dagger} \, \psi_{m-1}+D(m)\psi_m = - \frac{E}{t} \, \psi_m \, , \, m \ne 1,L, \notag \\
&R \, \psi_{2} + R^{\dagger} \, e^{-i q \theta \check\sigma_z} \psi_{L}+D(1)\psi_1 = - \frac{E}{t} \, \psi_1 , \notag \\
&R \, e^{i q \theta \check\sigma_z} \psi_{1} + R^{\dagger} \, \psi_{L-1}+D(L) \psi_L = - \frac{E}{t} \, \psi_L , \label{harper2}
\end{align}
where $\psi_m$ is a two-component spinor, $R= \bigl ( \cos(2\pi \gamma) \, \mathbb{I}+ i \sin( 2\pi \gamma) \check\sigma_x \bigr )$ and $D(m)$ is defined in Eq. \ref{Dmat}.

The topological invariant associated to the $r^{\txt{th}}$ bulk energy gap, $\txt{SChN} (r)$, is computed on a discrete Brillouin zone according to the expression \be\label{Z2invariant}
 \txt{SChN} (r)=\frac{i}{2 \pi} \sum_{\theta, k_y} \mathcal{F} (\Psi (\theta, k_y)) .
 \ee
Here $\mathcal{F} (\Psi)$ represents the non-Abelian Berry's curvature~\cite{Fukui2007} associated to the multiplet $\Psi= (\phi_1, \phi_2, \dots , \phi_N)$, where $\phi_{\lambda}$ denotes the single-particle wave-functions associated to the energy bands $E_{\lambda}$ lying below the $r^{\txt{th}}$ bulk gap.  The NI phase [i.e. $I_{Z_2}=0$] is reached in the $r^{\txt{th}}$ bulk gap when $\txt{SChN}(r)=0 \, \txt{mod} 4$, while this gap corresponds to the QSH phase [i.e. $I_{Z_2}=1$] when $\txt{SChN}(r)= \pm 2 \, \txt{mod} 4$. \\

In the uncoupled regime $\gamma=0$, our setup displays a double quantum Hall system in which particles with opposite spins are coupled to opposite magnetic fields, but not to each other. Such a system exhibits TR-symmetry, and therefore, bands come in time-reversed pairs characterized by  opposite Chern numbers: $\txt{ChN}_{\uparrow}= - \txt{ChN}_{\downarrow}$ . The net Chern number $\txt{ChN}=\txt{ChN}_{\uparrow}+ \txt{ChN}_{\downarrow}$ which accounts for the charge transport is zero, but the difference $\txt{SChN}=\txt{ChN}_{\uparrow}- \txt{ChN}_{\downarrow}$ is finite and defines a quantized spin Hall conductivity. The successive energy bulk gaps shown in Fig. ~\ref{FIG2} are associated to the Chern numbers [from top to bottom]  $\txt{ChN}_{\uparrow}=\{ -1, -2, 2, 1\}$ in agreement with the computed $Z_2$ indices  $I_{Z_2}= \{ 1 , 0 , 0 , 1 \}$. However, bulk gaps associated to even Chern numbers lead to a vanishing spin Hall conductivity, as such a system has even pairs of edge states. Therefore, in systems with TR symmetry, $Z_2$ invariants are the key to distinguish QSH and NI phases.


\newpage

\begin{acknowledgments} 
N.G. thanks the F.R.S-F.N.R.S (Belgium) for financial support. M.~L. aknowledges the Grants of Spanish MINCIN (FIS2008-00784 and QOIT), EU (NAMEQUAM), ERC (QUAGATUA) and of  Humboldt Foundation. A.~B. and M.-A. M.-D. thank the Spanish MICINN grant FIS2009-10061, CAM research consortium QUITEMAD, European FET-7 grant PICC, UCM-BS grant GICC-910758 and FPU MEC grant. I.~S. and P.~N. are supported by the grant N00014-09-1-1025A by the Office of
Naval Research, and the grant 70NANB7H6138, Am 001 by the National Institute
of Standards and Technology. I.~B.~S. was supported by the NSF through the JQI Physics Frontier
Center.  Correspondence and requests for materials should be addressed to N.~G.
\end{acknowledgments}







\begin{thebibliography}{99}

\newpage

\section*{References}

\bibitem{HasanKane} M. Z. Hasan $\&$ Kane, C. L. Topological Insulators, arXiv:1002.3895vl

\bibitem{Kitaev} Kitaev, A. Periodic table for topological insulators and superconductors.  {\it AIP Conf. Proc.} {\bf 1134,} 22 (2009).


\bibitem{Kane2006}
Kane, C. L.  $\&$ Mele, E. J.  {A new spin on the insulating state.} { \it  Science} {\bf 314,} 1692 (2006).

\bibitem{Kane2005bis}
Kane, C. L.  $\&$ Mele, E. J. {Quantum spin Hall effect in graphene.} {\it  Phys. Rev. Lett.}  \textbf{95,} 226801 (2005).

\bibitem{Bernevig2006}
Bernevig, B. A. $\&$ Zhang, S.-C.  {Quantum spin Hall effect.} {\it  Phys. Rev. Lett.}  \textbf{96,} 106802 (2006).


\bibitem{Bernevig2006bis}
Bernevig, B. A.,  Hughes, T. L.  $\&$ Zhang, S.-C. Quantum spin Hall effect and topological phase transition in HgTe quantum wells.
\emph{Science} {\bf 314,} 1757 (2006).


\bibitem{Fu2007}
Fu, L., Kane, C. L. $\&$ Mele, E. J. {Topological insulators in three dimensions}.
\emph{Phys.Rev. Lett.} {\bf 98,} 106803 (2007).

\bibitem{Konig2007}
K\"{o}nig, M., et al. {Quantum spin Hall insulator state in HeTe quantum wells}.
\emph{Science} {\bf 318,} 766 (2007).


\bibitem{Chen2009b}
Chen, Y. L. et al. {Experimental realization of a three-dimensional topological insulator, Bi$_2$Te$_3$}.
\emph{Science} {\bf 325,} 178 (2009).



\bibitem{Klitzing1980}
von Klitzing, K., Dorda, G., and Pepper, M.  {New method for high-accuracy determination of the fine-structure constant based on quantized Hall resistance}. \emph{Phys. Rev. Lett.}  \textbf{45,} 494 (1980).

\bibitem{Tsui1982} Tsui, D. C., Stormer, H. L., and Gossard, A. C. {Two-dimensional magnetotransport in the extreme quantum limit}. \emph{Phys. Rev. Lett.}  \textbf{48,} 1559 (1982).


\bibitem{Hatsugai1993}
Hatsugai, Y. {Edge states in the integer quantum Hall effect and the Riemann surface of the Bloch function}. \emph{Phys. Rev. B} {\bf 48,} 11851 (1993).


\bibitem{Kane2005}
Kane C. L. $\&$ Mele, E. J.  {$Z_2$ topological order and the quantum spin Hall effect}. \emph{Phys. Rev. Lett.}  \textbf{95,} 146802 (2005).


\bibitem{edge}
Wu, C.,  Bernevig, B. A. $\&$  Zhang, S.-C. {Helical liquid and the edge of quantum spin Hall systems}. \emph{Phys. Rev. Lett.} {\bf 96,} 106401 (2006).

\bibitem{edgebis}
Xu, C.  $\&$ Moore, J.  {Stability of the quantum spin Hall effect: Effects of interactions, disorder, and $Z_2$ topology}. \emph{Phys. Rev. B.} {\bf 73,} 045322 (2006).


\bibitem{Lewenstein2007}
Lewenstein, M. et al.  {Ultracold atomic gases in optical lattices: Mimicking condensed matter physics and beyond}. \emph{ Adv. Phys.} {\bf 56,} 243 (2007).


\bibitem{Bloch2008}
Bloch, I.,  Dalibard, J.  $\&$  Zwerger, W. {Many-body physics with ultracold gases}.  \emph{Rev. Mod. Phys.} {\bf 80,} 885 (2008).

\bibitem{Lin2009}
Y.-J. Lin, et al. {Bose-Einstein condensate in a uniform light-induced vector potential.} \emph{Phys. Rev. Lett.} \textbf{102,} 130401 (2009).

\bibitem{Lin2009bis}
Lin, Y.-J., Compton, R. L.,  Jiménez-García, K.,   Porto, J. V. $\&$ Spielman, I. B.  {Synthetic magnetic fields for ultracold neutral atoms.} \emph{Nature} {\bf 462,} 628 (2009).

\bibitem{Jaksch2003}
Jaksch, D.  $\&$ Zoller, P.  {Creation of effective magnetic fields in optical lattices: the Hofstadter butterfly for cold neutral atoms.} \emph{New J. Phys.} {\bf 5,} 56 (2003).

\bibitem{mueller}
Mueller, E. J.  {Artificial electromagnetism for neutral atoms: Escher staircase and Laughlin liquids}. \emph{Phys. Rev. A.} {\bf 70,} 041603(R) (2004).

\bibitem{Sorensen2004}
S{\o}rensen,  A. S.,  Demler, E.  $\&$ Lukin, M. D.  {Fractional quantum Hall states of atoms in optical lattices}. \emph{Phys. Rev. Lett.} \textbf{94,} 086803 (2004).

\bibitem{spielman}
Spielman, I. B.  {Raman processes and effective gauge potentials}. \emph{Phys. Rev. A.} \textbf{79,} 063613 (2009).

\bibitem{Gerbier2009} Gerbier, F.  $\&$ Dalibard, J. {Gauge fields for ultracold atoms in optical superlattices}. To appear in \emph{New J. Phys.} arXiv:0910.4606


\bibitem{Osterloh2005}
Osterloh, K., Baig, M.,  Santos, L., Zoller, P. $\&$ Lewenstein, M. {Cold atoms in non-Abelian gauge potentials: From the Hofstadter ``Moth" to lattice gauge theory}. \emph{Phys. Rev. Lett.} \textbf{95,} 010403 (2005).

\bibitem{Ruseckas2005}
Ruseckas, J., Juzeliunas, G.,  \"Ohberg, P. $\&$ Fleischhauer, M.  {Non-Abelian gauge potentials for ultracold atoms with degenerate dark states}. \emph{Phys. Rev. Lett.} \textbf{95,} 010404 (2005).


\bibitem{maciejarxiv} Sanchez-Palencia, L.  $\&$ Lewenstein, M. {Disordered quantum gases under control}. \emph{Nature Physics}  \textbf{6,} 87 (2010).






\bibitem{Stanescu2009}
Stanescu, T. D. et al.  {Topological Insulators and Metals in Atomic Optical Lattices}. \emph{Phys. Rev. A}  \textbf{79,} 053639 (2009).


\bibitem{Stanescuarxiv}  Stanescu, T. D., Galitski, V. $\&$ Das Sarma, S. Topological states in two-dimensional optical lattices. arXiv:0912.3559




\bibitem{Goldman2009}
Goldman, N., Kubasiak, A.,  Gaspard, P. $\&$ Lewenstein, M.  {Ultracold atomic gases in non-Abelian gauge potentials: The case of constant Wilson loop}. \emph{Phys. Rev. A.} {\bf 79,} 023624 (2009).


\bibitem{Goldman2009bis}
Goldman, N.  et al. {Non-Abelian optical lattices: Anomalous quantum Hall effect and Dirac fermions}. \emph{Phys. Rev. Lett.} {\bf 103,} 035301 (2009).



\bibitem{Satija1}
Satija, I.,  Dakin, D. C. $\&$  Clark, C. W. {Metal-insulator transition revisited for cold atoms in non-Abelian gauge potentials}. \emph{Phys Rev lett.} {\bf 97,} 216401, (2006).


\bibitem{Satija2}
Satija, I.,   Dakin, D. C., Vaishnav, J. Y. $\&$ Clark, C. W. { Two-dimensional electron gas with cold atoms in non-Abelian gauge potentials}.  {\it Phys Rev A} {\bf 77,} 043410, (2008).

\bibitem{Trinker2008}
Trinker, M., Groth, S., Haslinger, S., Manz, S., Betz, T., Schneider, S., Bar-Joseph, I., Schumm, T., $\&$ Schmiedmayer, J. { Multilayer atom chips for versatile atom micromanipulation}.  {\it Appl. Phys. Lett.} {\bf 92,} 254102, (2008).

\bibitem{Zhu2006}  Zhu, S.-L.,  Fu, H., Wu, C.-J.,  Zhang, S.-C. $\&$  Duan,  L.-M. {Spin Hall effects for cold atoms in a light-induced gauge potential}. {\it Phys. Rev. Lett.} \textbf{97,} 240401 (2006).

\bibitem{oh} Liu, X.-J., Liu, X., Kwek, L. C. $\&$  Oh, C. H. { Spin Hall effect in atoms}. {\it Phys. Rev. Lett.} {\bf 98,} 026602 (2007).


 \bibitem{Hofstadter1976} Hofstadter, D. {Energy levels and wave functions of Bloch electrons in rational and irrational magnetic fields}. {\it Phys. Rev. B} {\bf 14,} 2239 (1976).

\bibitem{biboutlook} Fu, L. $\&$  Kane, C. L. {Superconducting proximity effect and Majorana fermions at the surface of a topological insulator}. {\it Phys. Rev. Lett.} {\bf 100,} 096407 (2008).


\bibitem{biboutlook2} Qi, X.-L., Hughes, T. L., Raghu, S.   $\&$  Zhang, S.-C. {Time-reversal-invariant topological superconductors and superfluids in two and three dimensions}. {\it Phys. Rev. Lett.} {\bf 102,} 187001 (2009).




 \bibitem{Thouless1982} 
 Thouless, D. J., Kohmoto, M., 
Nightingale, M. P. $\&$ den Nijs, M.  {Quantized Hall conductance in a two-dimensional periodic potential}. {\it Phys. Rev. Lett.} {\bf 49,} 405 (1982).


 \bibitem{Niu1985} 
 Niu, Q. Thouless, D. J. $\&$ Wu, Y.-S.  {Quantized Hall conductance as a topological invariant}. {\it Phys. Rev. B} {\bf 31,} 3372 (1985).

 \bibitem{Kohmoto1985} Kohmoto, M. {Topological invariant and the quantization of the Hall conductance}. {\it Ann. Phys.} {\bf 160,} 343 (1985).


\bibitem{Sheng2006}  Sheng, D. N., Weng, Z. Y., Sheng, L. $\&$ Haldane, F. D. M.  {Quantum spin-Hall effect and topologically invariant Chern numbers}. {\it Phys. Rev. Lett.} {\bf 97,} 036808 (2006).

\bibitem{Qi2006} Qi, X.-L., Wu, Y.-S. $\&$ Zhang, S.-C. {General theorem relating the bulk topological number to edge states in two-dimensional insulators}. {\it Phys. Rev. B} {\bf 74,} 045125 (2006).


\bibitem{Fukui2007} Fukui, T. $\&$  Hatsugai, Y. {Topological aspects of the quantum spin-Hall effect in graphene: $Z_2$ topological order and spin Chern number}. {\it Phys. Rev. B} {\bf 75,} 121403(R) (2007).





\bibitem{Dao2007} Dao, T.-L., Georges, A., Dalibard, J., Salomon, C. $\&$ Carusotto, I. {Measuring the one-particle excitations of ultracold fermionic atoms by stimulated Raman spectroscopy}. {\it Phys. Rev. Lett.} {\bf 98,} 240402 (2007).

\bibitem{Stamper1999} Stamper-Kurn, D. M. et al. {Excitation of phonons in a Bose-Einstein condensate by light scattering}. {\it Phys. Rev. Lett.} {\bf 83,} 2876 (1999).

\bibitem{Liu2010} Liu, X.-J., Wu, C. $\&$ Sinova, J. {Quantum anomalous Hall effect with cold atoms trapped in a square lattice}. {\it Phys. Rev. A} {\bf 81,} 033622 (2010).

\bibitem{Umucalilar2008} Umucalilar, R. O.,  Zhai, H.  $\&$ Oktel, M. \"O. Trapped Fermi Gases in Rotating Optical Lattices: Realization and Detection of the Topological Hofstadter Insulator. {\it Phys. Rev. Lett.} {\bf 100,} 070402 (2008).


\bibitem{Haldane1988} Haldane, F. D. M. Model for a quantum Hall effect without Landau levels: Condensed-matter realization of the ``parity anomaly''. {\it Phys. Rev. Lett.} {\bf 61,} 2015 (1988).



\end{thebibliography}
\end{document}